\documentclass[journal]{IEEEtran}

\usepackage[utf8]{inputenc}
\usepackage{color}
\usepackage[pdftex]{graphicx}
\usepackage[caption=false,font=footnotesize]{subfig}
\usepackage{float}
\usepackage{multirow}
\usepackage[colorinlistoftodos]{todonotes}

\begin{document}

\title{Video Multimethod Assessment Fusion (VMAF) on 360VR contents}

\author{Marta~Orduna, C\'esar~D\'iaz, Lara~Mu\~noz, Pablo~P\'erez, Ignacio~Benito, and~Narciso~Garc\'ia\thanks{Manuscript received xxxx, 2018; revised xxxx, 2018.}\thanks{M. Orduna, C. D\'iaz, L. Mu\~noz and N. Garc\'ia are with the Grupo de Tratamiento
de Im\'agenes, Information Processing and Telecommunications Center and ETSI Telecomunicaci\'on, Universidad Polit\'ecnica de Madrid, 28040
Madrid, Spain (e-mail: moc@gti.ssr.upm.es, cdm@gti.ssr.upm.es, lms@gti.ssr.upm.es, narciso@gti.ssr.upm.es)}\thanks{P. P\'erez and I. Benito are with Nokia Bell Labs, Mar\'ia Tubau 9,
28050 Madrid, Spain (e-mail: pablo.perez@nokia-bell-labs.com, ignacio.benito\_frontelo@nokia-bell-labs.com)}\thanks{This work has been partially supported by the Ministerio de Ciencia, Innovación y Universidades (AEI/FEDER) of the Spanish Government under project TEC2016-75981 (IVME) and the Spanish Administration agency CDTI under project IDI-20170739 (AAVP).}}

\markboth{IEEE Transactions on Multimedia}%
{Design and implementation of QoE analysis of Super Multiview Video with Head-Mounted Devices}

\maketitle

\begin{abstract}
This paper describes the subjective experiments and subsequent analysis carried out to validate the application of one of the most robust and influential video quality metrics, Video Multimethod Assessment Fusion (VMAF), to 360VR contents. VMAF is a full reference metric initially designed to work with traditional 2D contents. Hence, at first, it cannot be assumed to be compatible with the particularities of the scenario where omnidirectional content is visualized using a Head-Mounted Display (HMD). Therefore, through a complete set of tests, we prove that this metric can be successfully used without any specific training or adjustments to obtain the quality of 360VR sequences actually perceived by users.  

\end{abstract}


\IEEEpeerreviewmaketitle

\section{Introduction}

\IEEEPARstart{V}{irtual} Reality (VR) applications try to provide an immersive experience to the user by creating a realistic-looking world, which can be static or responsive to the user's actions~\cite{burdea2003virtual}. Among the available sensory feedbacks, visual information is clearly the most important one to help the perception of being physically present in a non-physical world~\cite{mirror}. However, the rendering of high quality video imposes critical technical restrictions. On the one hand, its synthesis demands important computational resources and, on the other hand, its transmission requires very high bit rates. While local computing power seems to be widely available, video delivery assuring the suppression of incompatible sensory input does not~\cite{schubert}. So, many VR applications have been restricted to operate with local video information, although synthesized video could be generated online from delivered abstract representations. Moreover, the synchronized presentations of all multimedia data streams should not be forgotten~\cite{yuan}.

Recently 360VR content has stemmed as one of the most relevant scenarios related to VR. Specifically, its visualization by a Head-Mounted Display (HMD) allows a 3 Degrees-of-Freedom (DoF) scenario, used by a wide variety of applications in very different areas like education, medicine, or entertainment. Although, different applications consider content locally hosted, leading edge proposals required content located elsewhere, either stored or live recorded, and streamed to the client whenever required. Adaptive Bit Rate (ABR) streaming techniques are widely used~\cite{abr}, but the delivery of omnidirectional content with an acceptable quality is still a challenge in this scenario due to the amount of resources required. Typically, contents with at least 4K resolution and 60~fps are required to provide good Quality of Experience (QoE), guaranteeing an immersive and engaging experience~\cite{sreedhar2016viewport, el2016streaming, highframerate}. However, these high requirements in terms of image resolution and encoding quality lead to very high bit rates when they are encoded and delivered in the same way as traditional 2D content. This fact presents a serious problem considering the bandwidth required to stream this kind of content~\cite{diaz2018viability}. 

Therefore, to relax these strict conditions, different approaches can be considered. First, the design of new quality ladders leading to different perceptible levels of quality in 360VR contents. Second, efficient delivery schemes that take advantage of the intrinsic characteristics and nature of 360VR visualization, in the form of HMDs. In particular, existing schemes are typically based on the fact that only a portion of the received 360VR sequence, called Field Of View (FOV), is viewed by the user, and the specific portion depends on his/her point of view with respect to the scene at that particular moment~\cite{willemsen2009effects, gaddam2016tiling}. Therefore, only the area that is viewed by the user needs to be provided with high quality, decreasing the required overall bit rate. Moreover, other approaches take into account the users' behavior assuming that users tend to look at certain orientations or elements in the scene with higher probability than others. In this case, the content is prepared considering saliency or attention maps, leading to a better use of the bit rate~\cite{aladagli2017predicting},~\cite{rai2017dataset}. Additionally, other proposals exploit the peculiarities of the type of projection used to map the spherical image onto before the encoding and transmission processes: equirectangular, cubemap, pyramidal, equiangular cubemap...~\cite{el2016streaming,corbillon2017viewport,brown2017bringing}. Indeed, each projection impacts in a different way the quality of the different areas of the omnidirectional image. These proposals then aim at smartly differentiating and handling the information, mostly spatially, in terms of coding and/or transmission, so as to provide QoE to users and save bit rate simultaneously.

All these approaches require a quality metric that offers reliable results in the sense that it should be able to capture the quality actually perceived by users when these strategies are put into practice with several targets: test the strategy itself and properly select and adjust the parameters that influence its performance. Thus, a significant effort has been made to adapt some of the most popular and useful quality metrics of the traditional 2D world to 360VR scenarios.

Indeed, there exist several works in the literature referring to modifications of the Peak Signal-to-Noise Ratio (PSNR) metric to fit the specific features of 360VR content. Specifically, Lakshman et al.~\cite{yu2015framework} proposed a method called Sphere based PSNR computation (S-PSNR) where the distorted frame is projected onto a sphere before computing its distortion. In this way, for each projected point on the sphere, the associated pixels in the plane domain are calculated to compute the PSNR. Based on the S-PSNR, other methods have targeted the approximation of the average quality over all possible user points of view which are related with different viewports, weighting the values obtained for a given viewport taking into account the probability of the users of looking in that direction. For instance, Sun et al.~\cite{wspsnr} proposed the use of the Weighted to Spherically PSNR (WS-PSNR) metric, where the weights assigned to an area decreases as this area gets away from the equator and closer to the poles. Similarly, Zakharchenko et al.~\cite{cpp} proposed the Craster Parabolic Projection PSNR (CPP-PSNR) metric, where the weights are assigned to different areas based on the craster parabolic projection. In contrast, Ghaznavi et al.~\cite{ghaznavi2016360} introduced the Uniformly Sampled Spherical PSNR (USS-PSNR) metric, where an uniform and equal weight sampling of the decoded video on the sphere is implemented. Hence, the sample density changes based on latitude and longitude. Anyhow, the main problem with this kind of metrics is that they still have the same problem as the original PSNR, they do not take into account any Human Visual System (HVS) characteristics.

With the aim of including subjective aspects in the way video quality is measured, a Multi-Scale SSIM (MS-SSIM) extension was proposed by Wang et al.~\cite{wang2003multiscale}. Starting from the comparison of the three traditional SSIM terms (luminance, contrast, and structure) between the original and the distorted sequences, this extension incorporated information regarding image details at different resolutions and several viewing conditions that some subsequent works have adjusted to be used with omnidirectional content. Specifically, the version by Corbillon et al.~\cite{corbillon2017viewport} use different encoded versions of the same viewport whereas the proposal by Tran et al.~\cite{tran2018study} is based on different encoded versions of the whole 360VR scene. Nevertheless, although the approximation of the perceived quality carried out by MS-SSIM is in general acceptable and outperforms the results of PSNR, the complexity of applying this index to omnidirectional contents of high resolutions complicates its use~\cite{tran2018study}.

Based on this overview, none of the modifications of traditional objective metrics offers useful enough evaluations in terms of reliability and resource consumption. For this reason, we have focused our work on the extension to omnidirectional video one of the most influential metrics used today for traditional contents: the Video Multimethod Assessment Fusion (VMAF) metric developed by Netflix~\cite{liu2013visual,lin2014fusion,vmafblog}. VMAF is a Full-Reference (FR) metric based on different elementary metrics combined by a machine-learning algorithm, offering a good prediction of the human quality perception~\cite{vmafblog}. Recent studies have validated its direct use on environments different from the one it was intended to without any specific training in this sense. Concretely, Rassol et al.~\cite{vmafvalidation4k} carried out subjective quality tests to validate the application of VMAF to traditional contents with 4K resolution, a resolution for which the metric is not trained, obtaining good results when trying to predict the VMAF score. Bampis et al.~\cite{bampis2017learning} used the dataset created for VMAF to implement their quality predictor and compare the results obtained by VMAF with other typical metrics. Likewise, Bampis et al.~\cite{vmafextensionSTVMAF} proposed the SpatioTemporal-VMAF (ST-VMAF), an extension to the VMAF metric consisting in expanding the analysis of temporal features in video sequences to enhance the metric results. The significantly good results provided by VMAF with different type of non-immersive contents and viewing conditions led to considering its application without making any specific adjustments to assess omnidirectional content, thus avoiding generating a large and rich specific 360VR video dataset, carrying out numerous subjective quality assessments and performing the corresponding training and testing stages. Hence, saving time and resources.

The rest of the paper is structured as follows. Section~\ref{sec:work_approach} describes in detail the approach we have taken to validate the use of VMAF to assess the quality of 360VR content. Section~\ref{sec:application_vmaf_360vr} introduces the first stage of our procedure. In Section~\ref{sec:validation_vmaf_360vr}, we present the preparation, carrying out and analysis of the subjective quality assessment to validate the use of VMAF for 360VR contents. At the end of this paper, Section~\ref{sec:conclusions} summarizes general conclusions.

\begin{figure*}
    \centering
    \subfloat[\textit{AbandonedBuilding}~\cite{databaseclips}.]
    {
        \includegraphics[width=1.7in]{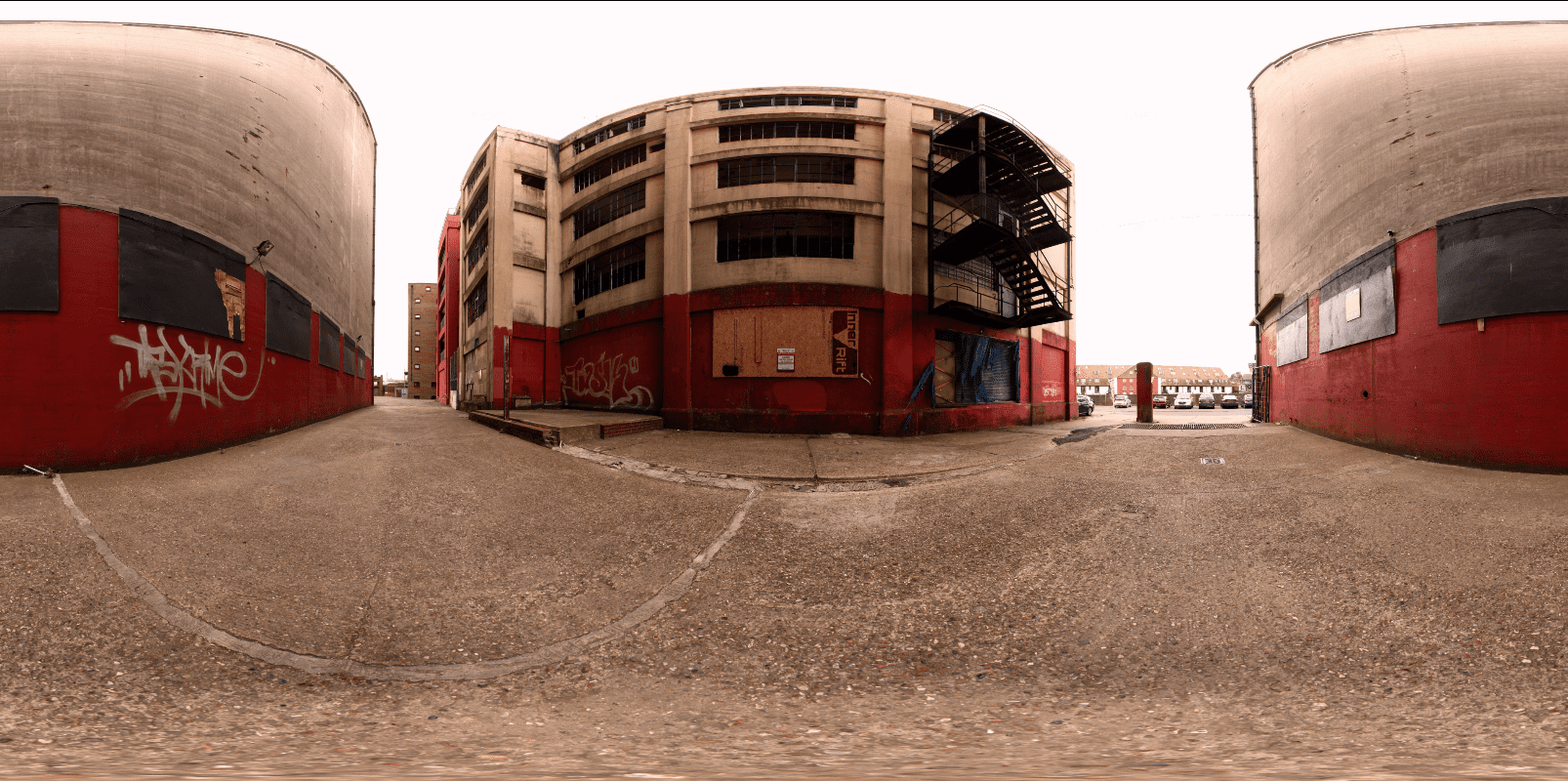}
    }
   \subfloat[\textit{Alaska}~\cite{databaseclips}.]
    {
        \includegraphics[width=1.7in]{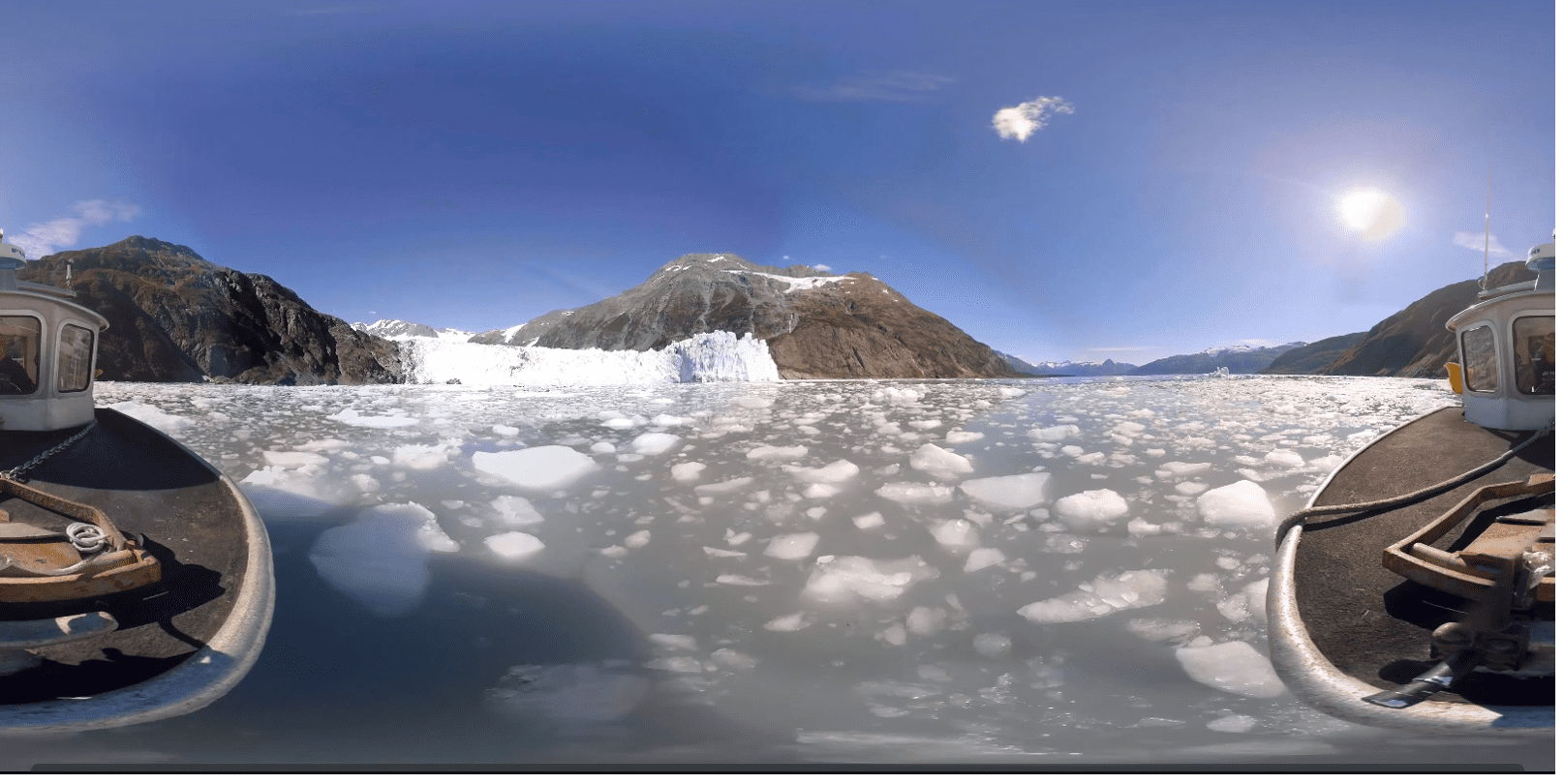}
    }
    \subfloat[\textit{Beach}~\cite{databaseclips}.]
    {
        \includegraphics[width=1.7in]{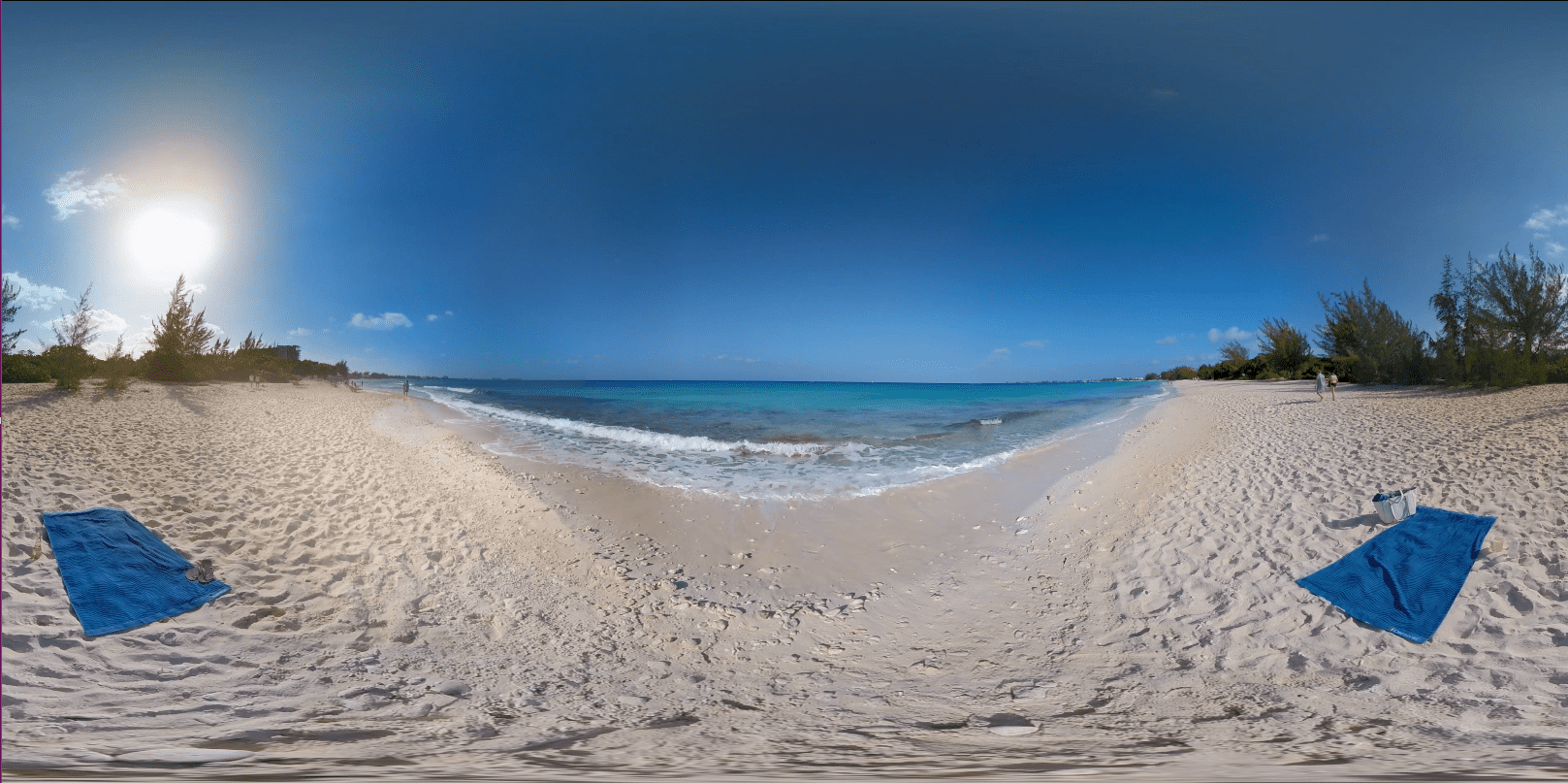}
    }
    \subfloat[\textit{CaribbeanVacation}~\cite{databaseclips}.]
    {
        \includegraphics[width=1.7in]{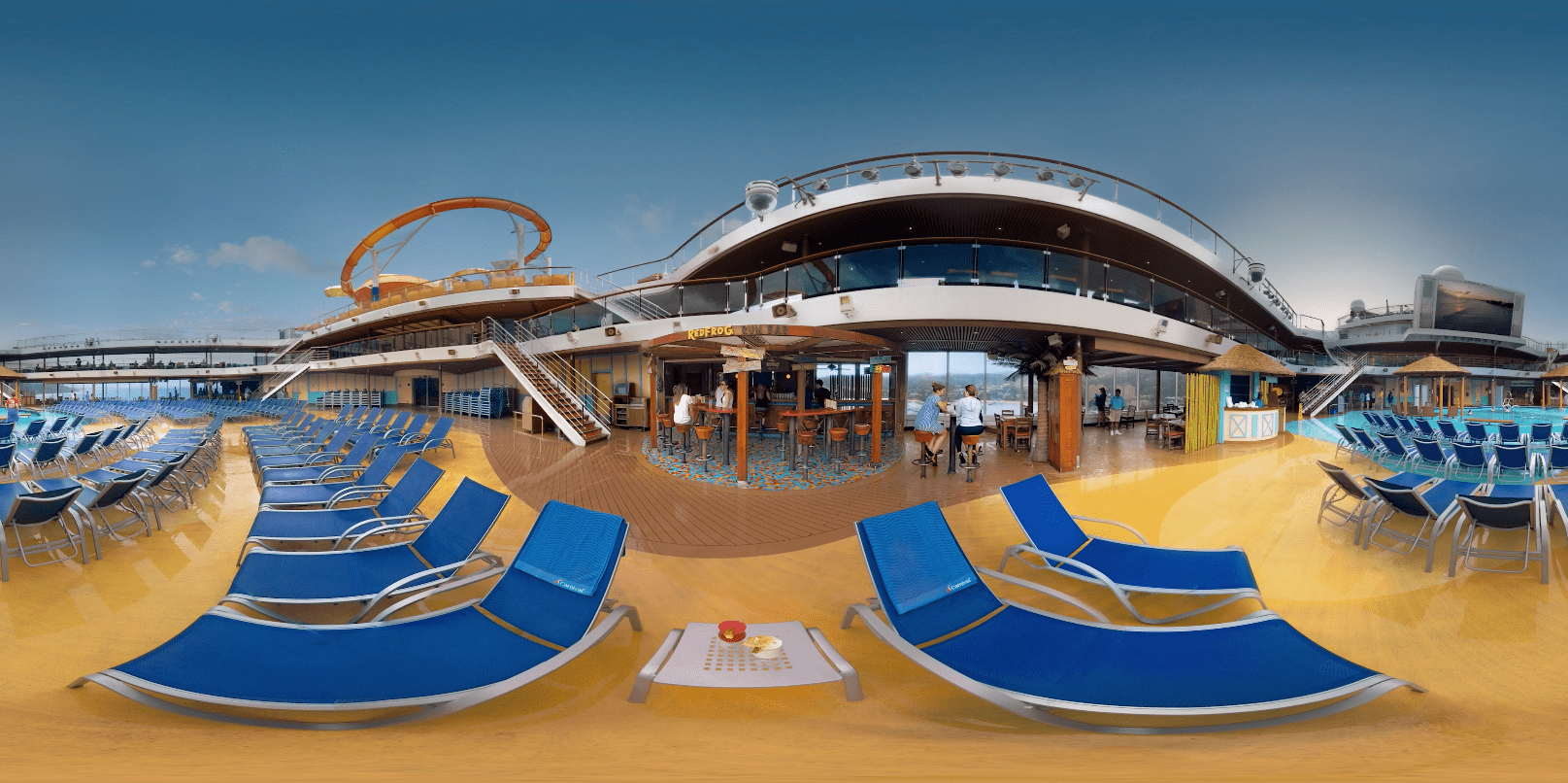}
    }
    \\
    \subfloat[\textit{FemaleBasket}~\cite{femaledataset}.]
    {
        \includegraphics[width=1.7in]{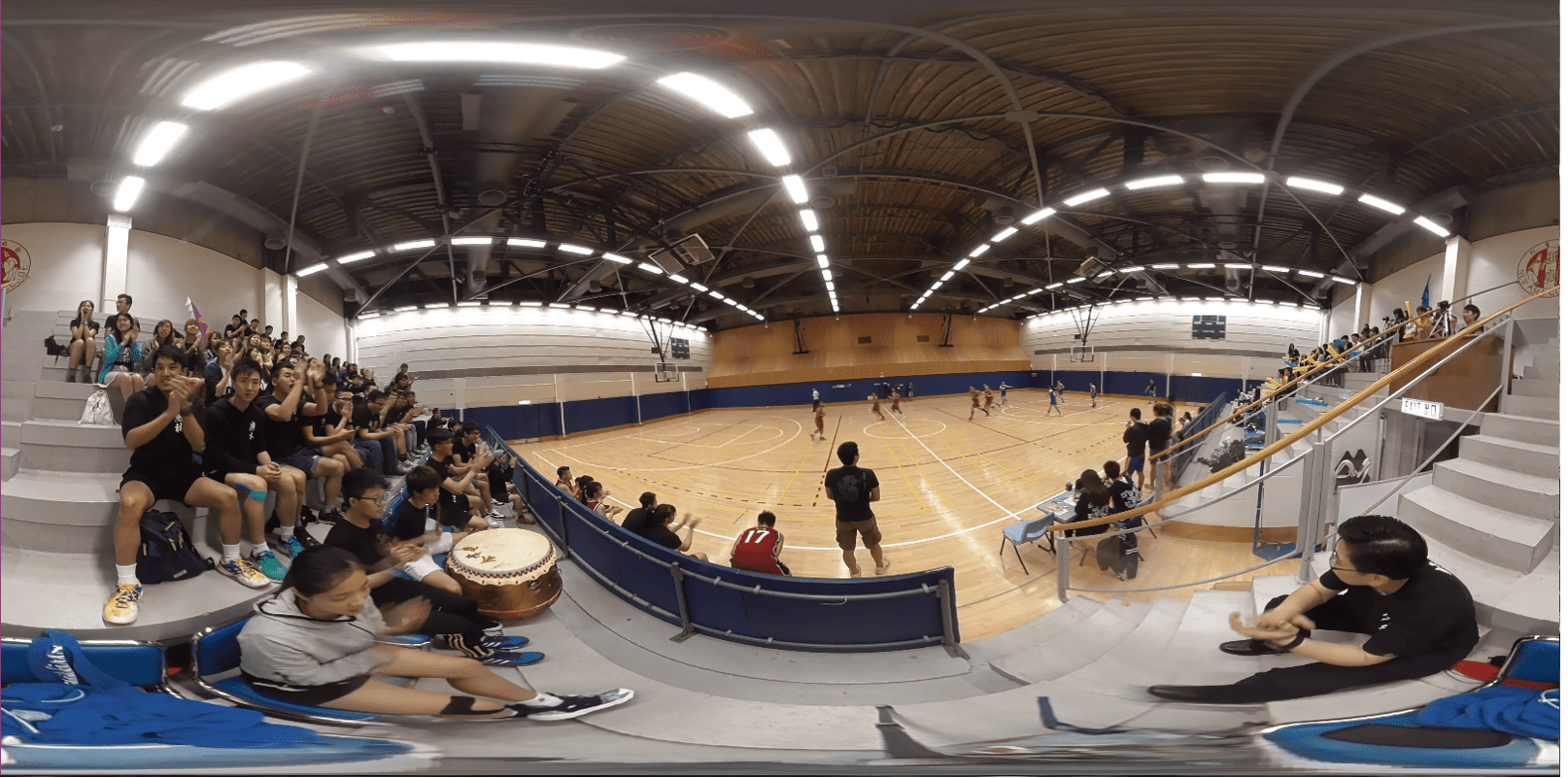}
    }
    \subfloat[\textit{Happyland}~\cite{databaseclips}.]
    {
        \includegraphics[width=1.7in]{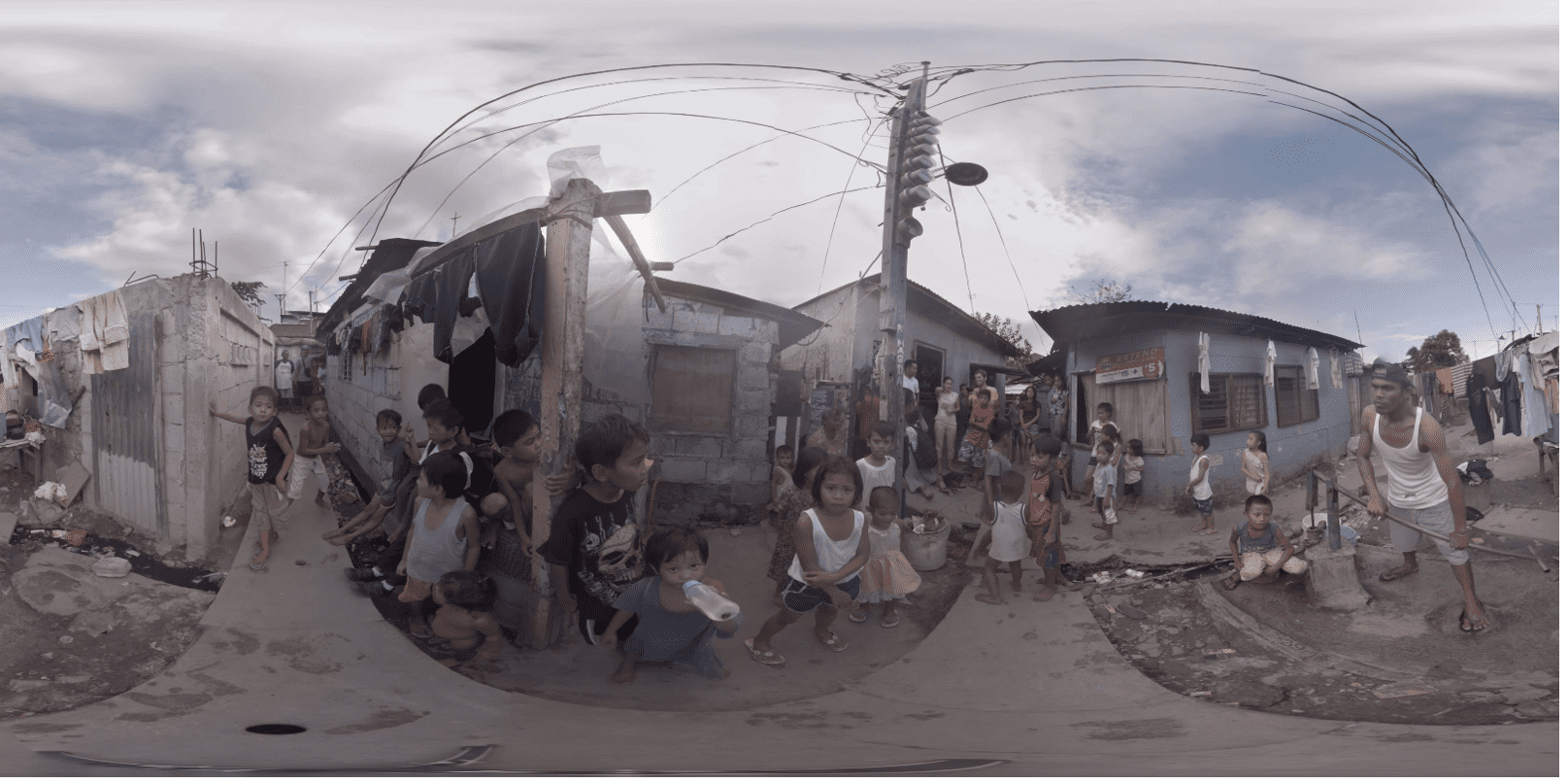}
    }
    \subfloat[\textit{Sunset}~\cite{databaseclips}.]
    {
        \includegraphics[width=1.7in]{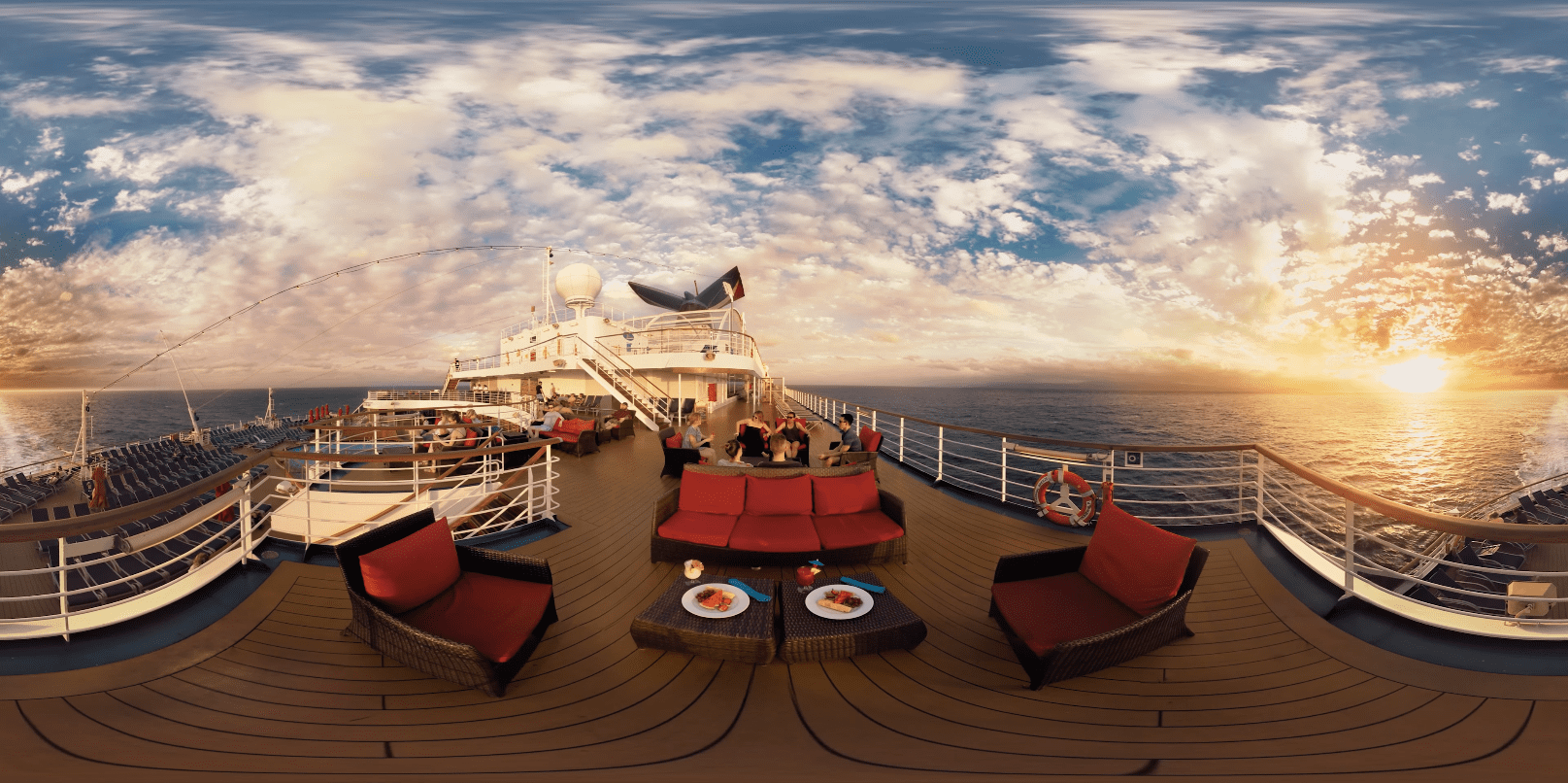}
    }
    \subfloat[\textit{Waterfall}~\cite{databaseclips}.]
    {
        \includegraphics[width=1.7in]{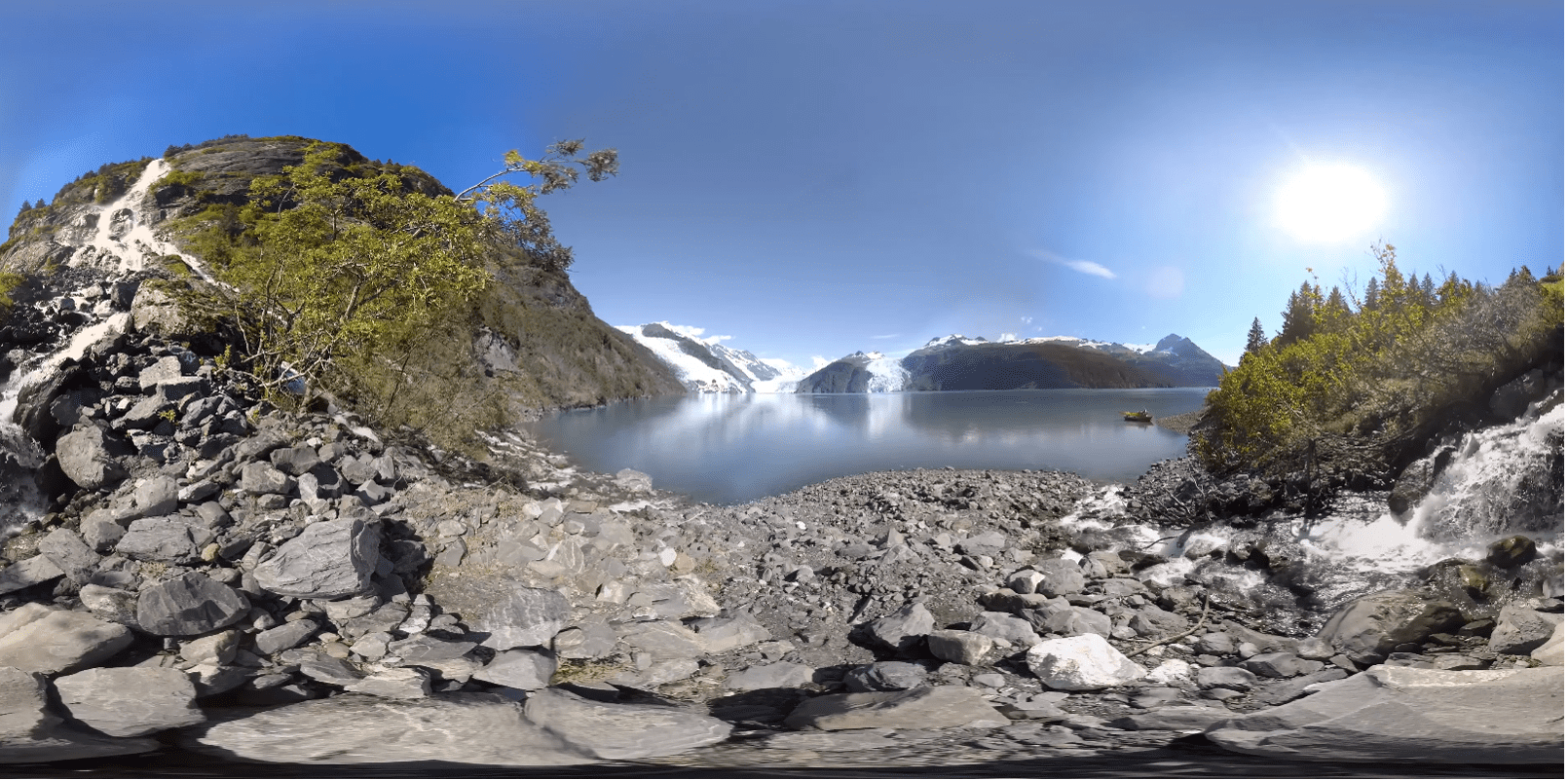}
    }
    \caption{Video sources screenshots.}
    \label{fig:frames_Example}
\end{figure*}

\section{Work approach}
\label{sec:work_approach}

The objective of this work is the validation of the direct application of the FR VMAF metric to omnidirectional content without any specific training or adaptations in this sense. To do so, we presume that there is a monotonic relationship between the well-known application of VMAF on traditional 2D contents and its proposed new application on 360VR contents. Therefore, the validation can be carried out on a reduced set of adequately selected values.

 The validation is performed in two steps. First, we encode a number of 360VR Source Sequences (SRCs) with constant Quantization Parameter (QP) covering the whole range of possible values. Later on, we simply apply the original VMAF metric to these Processed Video Sequences (PVS's) to obtain the variation of the score with the encoding parameter. It is posed in this way considering the high impact of QP in the QoE of the users. Indeed, in general, the higher the QP value is, the less detail is retained in the encoded image. This usually translates into lower QoE but also lower bit rate usage~\cite{qp}. Secondly, we verify through subjective tests that the users' perception fits with minimum discrepancy the VMAF scores obtained in the first step. Instead of performing a sweep over the whole range of QP values, i.e., showing to the user all the PVS's created in the first step, to search for Just-Noticeable Differences (JND), we present a subset of them and analyze their responses in terms of average rate and tendency. We do so assuming that the VMAF-vs-QP curve is monotonically decreasing by the nature of the encoding. This fact enables the possibility of adjusting it with a finite number of key operating points. These points correspond to anchor VMAF scores in the curve for all the used contents. 

We have focused on the equirectangular projection for our study, as it is the one most commonly used today. In addition, attention maps obtained from ordinary 360VR content visualization sessions show that users tend to look at the areas near the equator with a higher probability~\cite{maugey2017saliency}. Since the distortion introduced by the equirectangular projection is far lower in these areas, the characteristics of the zone are closer to those of traditional 2D contents. Thus, we can assume that a robust metric designed for 2D content can offer acceptable results for common 360VR content with homogeneous encoding.

The following sections describe in depth both steps.

\section{Application of VMAF to 360VR contents}
\label{sec:application_vmaf_360vr}

Here, we present and show the reasons why we use this process through which we obtain the reference VMAF-vs-QP curve for 360VR contents. It is divided into two main parts: the test material subsection, where the created database and the main features of the SRCs are presented, and the experimental results subsection, where the VMAF scores are presented and analyzed.

\begin{figure}[t]
    \centering
    \includegraphics[width=\columnwidth]{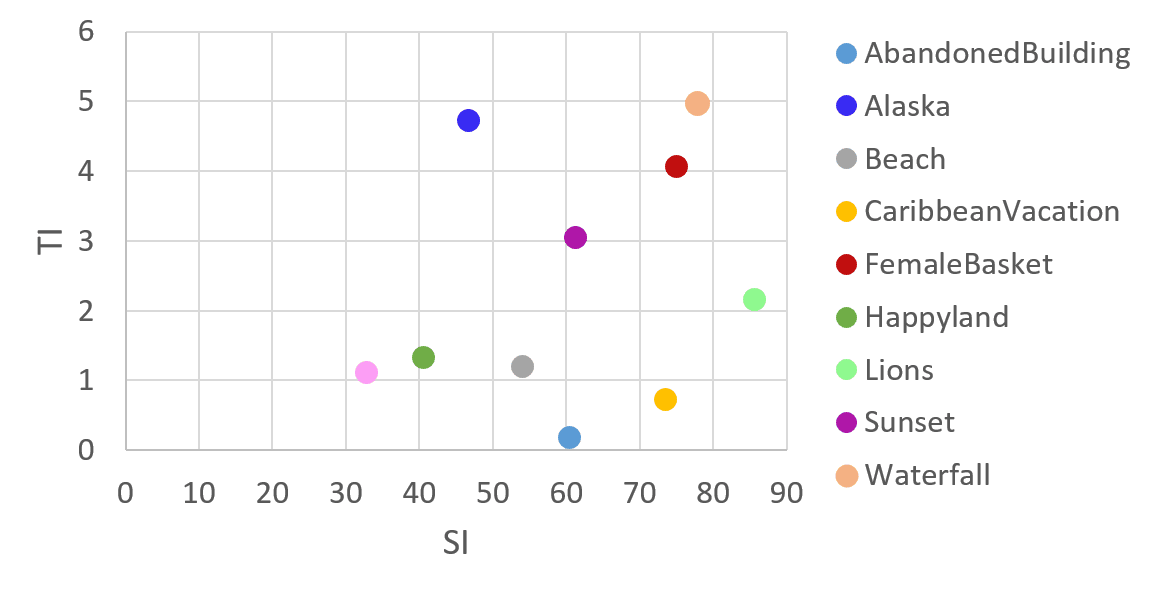}
    \caption{Spatial and Temporal Information indicators for all contents.\label{fig:SI_TI_DATA}}
\end{figure}

\subsection{Test material}

The first step for this analysis was to prepare a wide range of 360VR contents selected with different features in terms of color, texture, camera motion, composition, and type of content in the scenes~\cite{machajdik2010affective}, in accordance to Recommendation ITU-R BT.500-13~\cite{bt2012methodology}. It is important to note the relevance of the dataset content selection because a varied material with an absence of defects must be assured to obtain stable results in this analysis and in the following subjective analysis (Section~\ref{sec:validation_vmaf_360vr}). Additionally, the selected clips should not present any relevant changes between frames, avoiding the need for an accurate temporal pooling mechanism. Furthermore, a minimum level of visual comfort should also be guaranteed. To that end, we did not consider clips that included abrupt movements in the scene, a poor stitching, or unbearable effects that could disturb subjects and affect their rates.

We used nine SRCs from as many immersive VR video sources in equirectangular format. Seven of them were obtained from a database made publicly available by the Virtual Human Interaction Lab from Stanford University~\cite{databaseclips} and one from the dataset for exploring user behaviors in VR spherical video streaming created by Wu et al.~\cite{femaledataset}. The last one came from a private source. Figure~\ref{fig:frames_Example} depicts descriptive screenshots of the first eight sequences. All nine clips had a duration of 10 seconds. This length is justified by previously conducted studies conclusions, where we detected that this is the average time that it usually takes users to properly explore a 360VR scene, that is, to find and check the anchor points in the 360 scene that he/she uses as reference to compare the quality between different versions. Furthermore, this duration matches the suggestions of Recommendation ITU-R BT.500-13~\cite{bt2012methodology}. Moreover, the original resolution of all the sequences is 4K (3840x1920), which was kept constant for all tested qualities throughout the experiment. As the original sources had different framerates, all clips were changed to 25~fps to build a homogeneous dataset. We next describe the main characteristics of the selected contents:
\renewcommand{\labelenumi}{\alph{enumi})}
\begin{enumerate}
    \item "AbandonedBuilding": it is mainly a static content with notable texture. The only motion in the scene is related to a moving curtain.
    \item "Alaska": this content's main feature relies on the motion of the camera, since it is on a sailing boat.
    \item "Beach": this content presents a typical beach landscape. The most relevant feature is the appearance of some titles, which can capture the user's attention.
    \item "CaribbeanVacation": this content shows a cruise with people in a cafeteria. A video is shown to the public, trying to capt the attention of the user.
    \item "FemaleBasket": this content presents a basketball game with people cheering.
    \item "Happyland": this content is characterized by the proximity of some children moving around the camera. 
    \item "Sunset": this content can be considered an exploratory content. The camera is on a sailing cruise. However, the camera motion is not perceptible because of the height of the cruise.
     \item "Waterfall": this content shows a landscape with a quite large waterfall that is rather close to the camera.
     \item "Lions": this content shows a lion moving very close around the camera.
\end{enumerate}

\begin{figure*}
    \centering
    \includegraphics[width=0.85\textwidth]{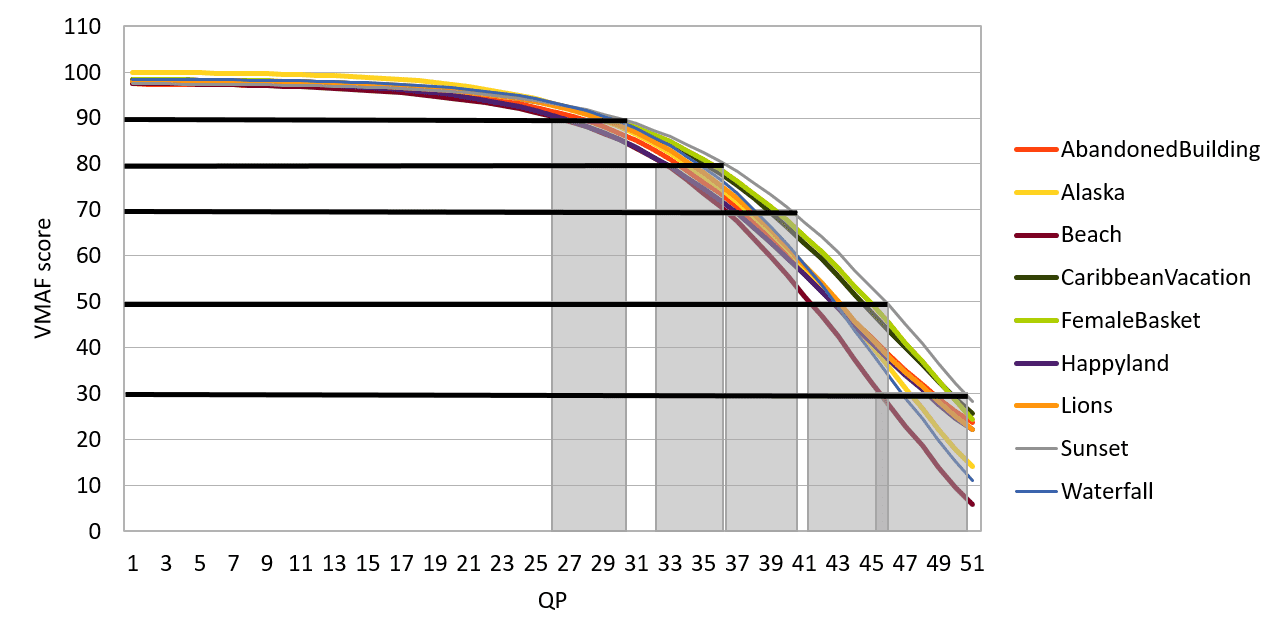}
    \caption{VMAF-vs-QP curve for all contents. The VMAF anchor values used for the validation are included.}
    \label{fig:vmaf_values}
\end{figure*}

All SRCs were characterized in terms of their spatial and temporal complexity, using the Spatial Information (SI) and Temporal Information (TI) indicators, respectively, as expressed in Recommendation ITU-T P.910~\cite{itutp910}. Their values are presented in Figure~\ref{fig:SI_TI_DATA}.

To obtain the full range of scores, all SRCs were encoded with ITU-T H.265/High Efficiency Video Coding (HEVC) using fixed QPs ranging from 1 to 51~\cite{ituh265}. As a result, we obtained 51 PVSs per SRC with bit rates ranging from 310~Mbps to 370~kbps. A summary of the created dataset is presented in Table~\ref{tab:dataset_qptest}. This set of 459 (51 times 9) sequences were the inputs to the VMAF computing algorithm.

\begin{table}[t]
\caption{Dataset characteristics\label{tab:dataset_qptest}}
\centering
\begin{tabular}{|l|l|}
\hline
Number of reference videos & 9 \\
\hline
Duration & 10~seconds \\
\hline
Encoding & H.265/HEVC \\
\hline
Resolution & 4K (3840x1920)  \\
\hline
Hypothetical Reference Circuits (HRCs)  & QP range (1-51)  \\
\hline
Framerate & 25~fps \\
\hline
\multicolumn{2}{|c|}{\textbf{Total number of videos: 459}} \\
\hline
\end{tabular}
\end{table}

\subsection{VMAF results\label{sec:objresults}}
In this subection, we present the results of computing VMAF metric to the whole set of PVSs. To that end, we used the VMAF Development Kit (VDK) that can be found available in a public repository~\cite{vmafgithub}. Particularly, we employed VDK version 1.3.3 and VMAF version 0.6.1. As has been justified previously, due to the absence of scene changes in the selected clips, the arithmetic mean was used as temporal pooling mechanism, since it is a representative value for those sequences.

Figure~\ref{fig:vmaf_values} shows the VMAF final scores for all contents in the whole range of tested QP values. It can be seen that the quality measured by this metric decreases monotonically with QP. Furthermore, the curve decreases slightly for the highest qualities (low QP values), more sharply for medium qualities (medium QP values), and dramatically for low qualities (high QP values). Besides, as already mentioned, the effect of changing the QP value varies with the characteristics of the content, resulting in a different VMAF curve for each of the SRCs.

\begin{figure*}[t]
    \centering
    \subfloat[\textit{MOS.}]
    {
        \includegraphics[width=0.85\textwidth]{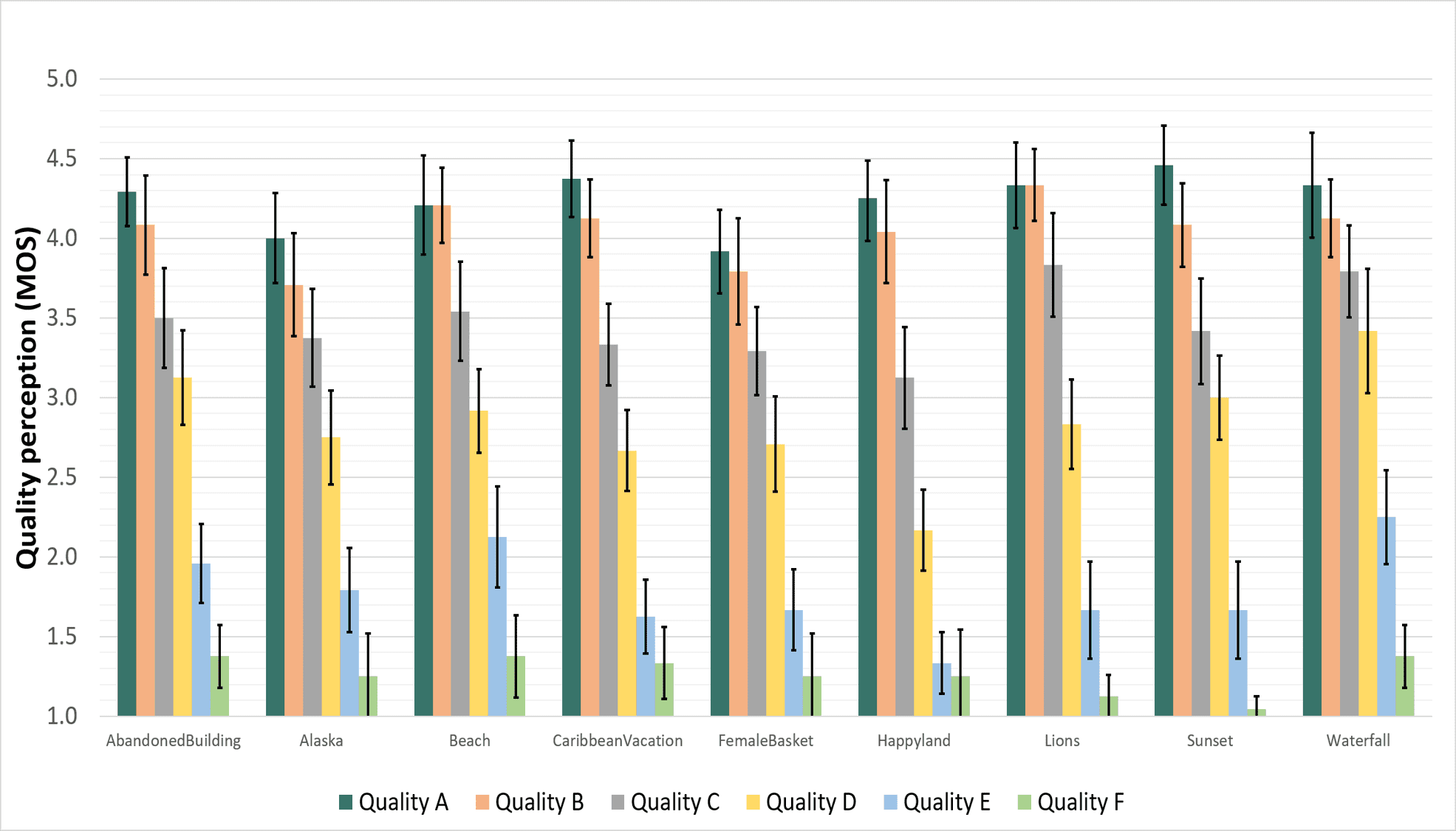}\label{fig:mos}
    }
   \\
   \subfloat[\textit{DMOS.}]
    {
        \includegraphics[width=0.85\textwidth]{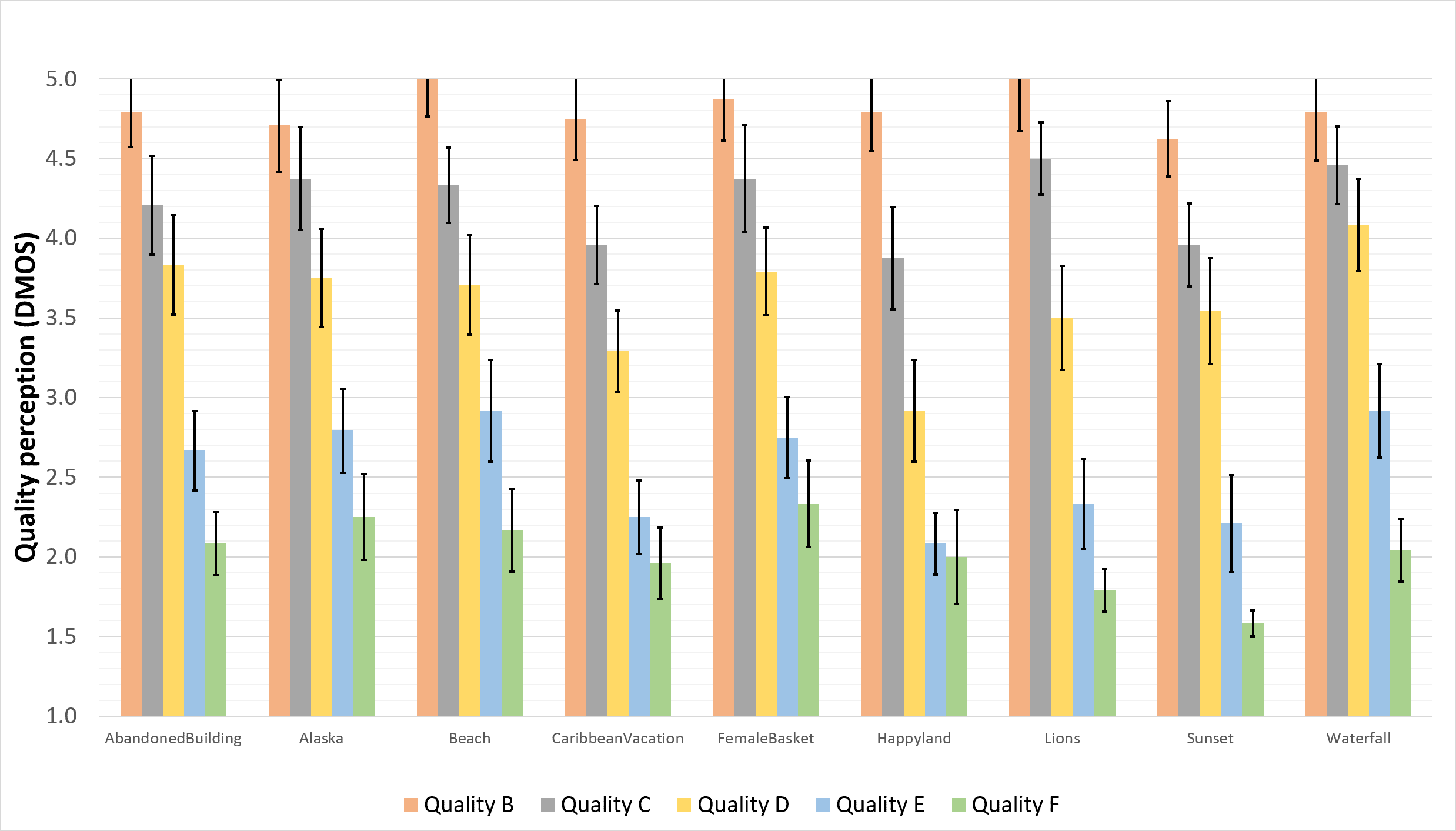}\label{fig:dmos}
    }
    \caption{MOS and DMOS for all contents and qualities, including CIs.}
\end{figure*}

\section{Validation of VMAF for 360VR contents through subjective quality assessment\label{sec:validation_vmaf_360vr}}
We describe in this section the subjective quality test conducted to validate the results obtained with VMAF. As mentioned above, VMAF is a metric prepared to work with traditional 2D contents. In this work, we evaluate to what extent it can be used with omnidirectional contents. To that end, we designed an experiment consisting in presenting a subset of the PVS's used in the previous step that are located closest to several strategic VMAF scores to a number of subjects. For each version, subjects were asked to evaluate the perceived quality. In this way, we obtained subjective quality rates for those strategic points along the QP range. These evaluations are used to check how close the given rates are from the objectively computed VMAF scores for 360VR contents.

Furthermore, it is first noteworthy mentioning that to date there are no official recommendations for subjective metrics to measure the QoE in 360VR scenarios, where HMDs are used as displays. Indeed, the first draft created by ITU-T Study group 12~\cite{G360VR} was published in early 2018 but the final document is still under development. In this way, the subjective assessment carried out in this work is based on the information obtained from recommendation documents related to traditional contents which have been highly tested: ITU-R BT.500-13~\cite{bt2012methodology}, ITU-T P.910~\cite{itutp910}, and ITU-T P.913~\cite{p913}.

\subsection{Test Material}
As mentioned, the test material for the subjective quality assessment is a subset of the PVS's generated and used in the previous step. In particular, we used the PVS's corresponding to six different quality levels: five distorted and one reference sequences. So, a total of 54 (six qualities, nine SRCs) are presented to each subject. Concretely, considering the VMAF curve in Figure~\ref{fig:vmaf_values}, the five distorted PVS's selected in the validation step are those closest to the following key VMAF scores:
\begin{itemize}
    \item VMAF equal to 90. This value is located where the curve begins to decrease slightly. 
    \item VMAF equal to 80 and 70. These values are located where the curve decreases more sharply. 
    \item VMAF equal to 50 and 30. These values are located where the curve decreases more dramatically.
\end{itemize}

Additionally, with respect to the reference sequences, on the one hand, we have no access to the original raw videos, but to encoded, and therefore degraded to some extent, sequences. On the other hand, references must comply with the same restrictions as the rest of sequences in the experiment, namely, that are encoded using a fixed uniform QP value. Therefore, we cannot directly use the available SRCs, but clips picked from the already generated PVS's database. So, for each content, we have selected a reference that scores higher than 90 in the VMAF scale, since the reference clip needs to offer the best quality presented to the user during the test, and which is encoded, when possible, with a similar bit rate to that of the original video. In this way, all selected sequences provide VMAF scores that range between 92 and 95.

The six qualities are denoted from A to F, where A is the reference (best quality version), and B to F are the five distorted versions associated with the VMAF scores 90, 80, 70, 50, and 30, respectively.

\subsection{Equipment}
The tests have been carried out in the smartphone Samsung Galaxy S8 with the last model of Samsung Gear VR glasses. This decision is based on the fact that consumer electronics devices are the most used for the 360VR content visualization application~\cite{device}.

\subsection{Environment}
The test area is set according to ITU-R BT.500-13~\cite{bt2012methodology}, creating an immersive space around the subject. In the set environment, we use a common HMD which only tracks the rotational movements~\cite{el2016streaming}. That is, it provides the 3 DoF that characterize this scenario. The selected location is in the middle of a room where the subject has no limitations to spin around.

The position is an important component of these subjective tests. For that, a swivel chair is used, as this kind of chairs allows subjects to move rather freely to see around them, facilitating the exploration of content. Naturally, the same chair is used for all subjects.

\subsection{Observers}
A total of 24 observers (8 females, 16 males) participated in this experiment. All of them with normal or corrected vision (glasses or contact lens are compatible with the equipment). The age of the subjects ranges from 21 to 36, with an average age of 26. Furthermore, the Pearson correlation between the data provided by each subject and the average of all resulted in that no subject was removed because of being considered an outlier~\cite{p913}. 

\subsection{Methodology}
 A Single-Stimulus (SS) method is applied in this experiment, specifically the ACR-HR (Absolute Category Rating with Hidden Reference)~\cite{itutp910}. In the conducted tests, there is no training session in terms of showing the expected maximum and minimum qualities to the subjects, because we want to observe the real absolute quality that they perceive. In this sense, it is possible that the best quality offered is not considered excellent by most subjects because of several factors: the specific features of the devices employed in the experiments, the network limitations, the quality of the original videos, and others. This effects can be later on considered or even partially cancelled in the subsequent analysis, thanks to the use of the hidden references. Indeed, with this method, a reference version of each content is randomly presented to subjects, without being paired with any distorted versions, and it is rated like any other~\cite{p913}. Later on, we can use the rates given to these hidden references to restrict  as much as possible the exogenous factors listed above during the analysis of the results. 
  
 The ACR-HR method uses the same five-level rating scale as the ACR method. According to Recommendation ITU-T P.910~\cite{itutp910}, the numbers may only optionally be displayed on the scale. Here, in our experiment, only the category ("Excellent", "Good", "Fair", "Poor" and "Bad") is displayed.

\subsection{Test session}
Subjects use a developed application program that allows for watching contents and rating them one after the other without having to remove his/her glasses or interact outside the 360VR environment~\cite{applicationtest}. This app then enables a more immersive and engaging experience for the subjects. Subjects are instructed at the beginning of the test session and guided if they have any problems with the app or the methodology during the test.

Each test session is composed of a total of 54 video clips (45 distorted and 9 reference videos) with a duration of 10~seconds each one. All videos are viewed by every subject. The duration of the whole test is around 15~minutes, assuming a period of approximately 5~seconds to vote each video clip. The voting period length is user-driven and so is not limited beforehand.
 
A different randomization of the PVS's is used for each session to reduce contextual effects. An observer can watch the same quality in two consecutive videos. However, subjects cannot watch the same clip with different qualities consecutively.

\begin{figure*}
    \centering
    \subfloat[\textit{AbandonedBuilding}.]
    {
        \includegraphics[width=0.4\textwidth]{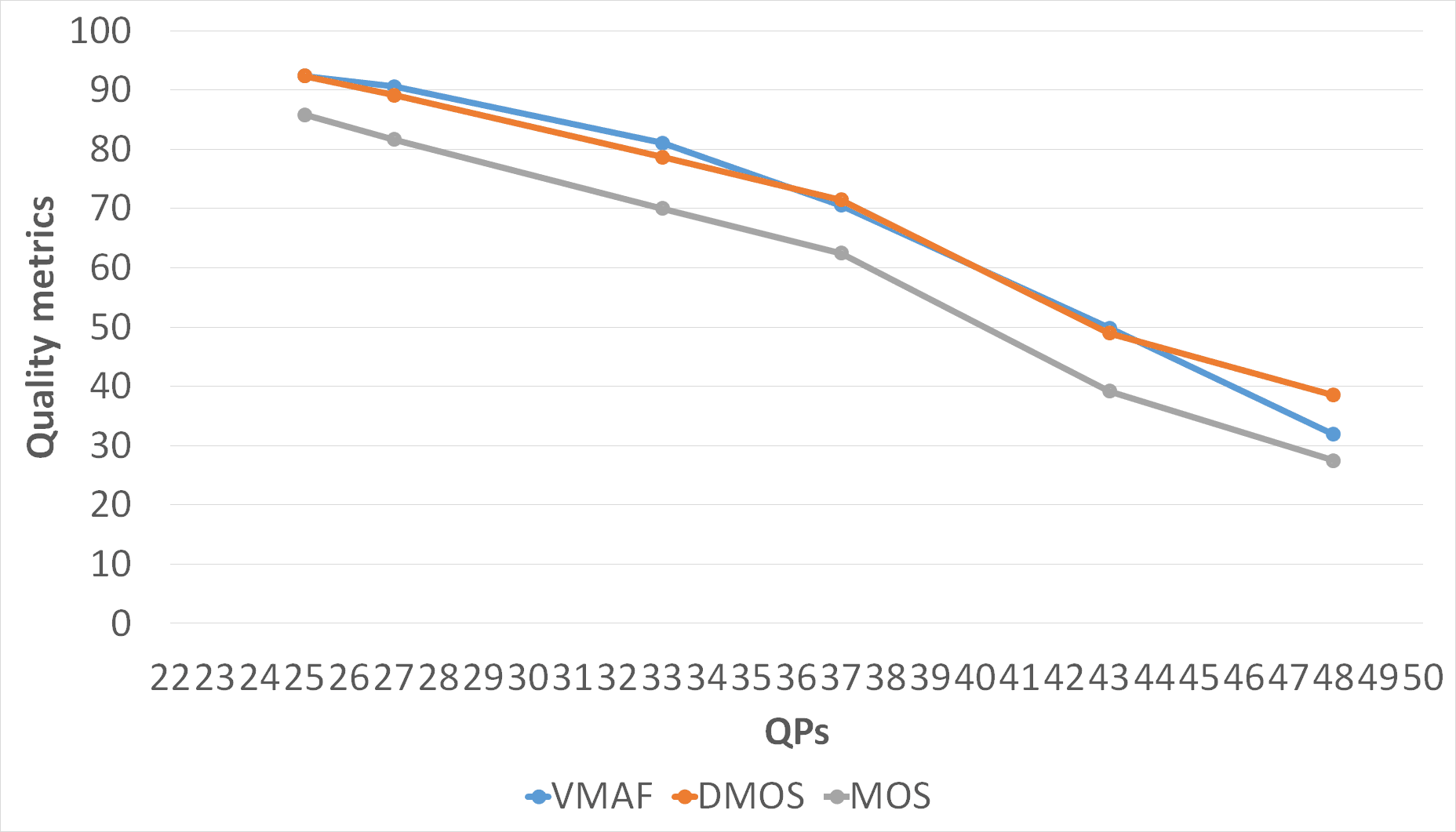}
    }
    \subfloat[\textit{Alaska}.]
    {
        \includegraphics[width=0.4\textwidth]{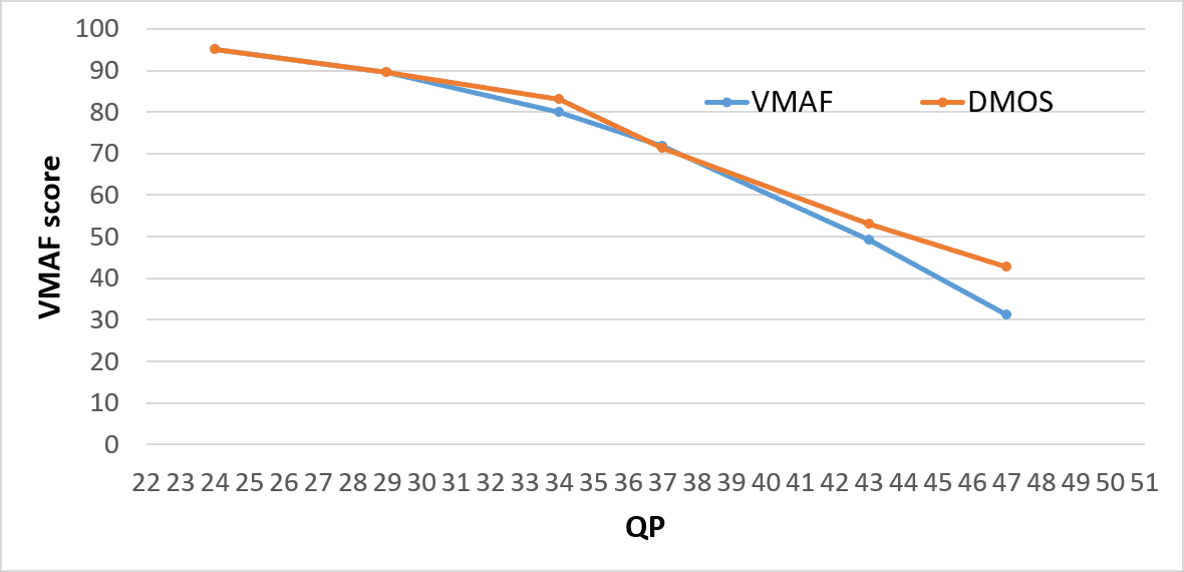}
    }
    \\
    
    \subfloat[\textit{Beach}.]
    {
        \includegraphics[width=0.4\textwidth]{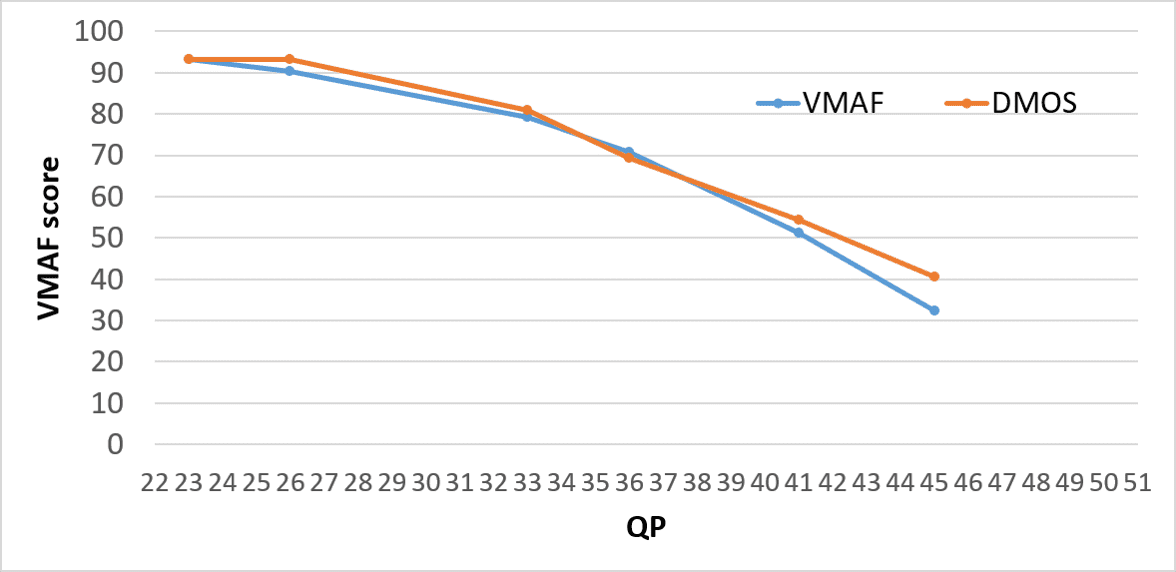}
    }
     \subfloat[\textit{CaribbeanVacation}.]
    {
        \includegraphics[width=0.4\textwidth]{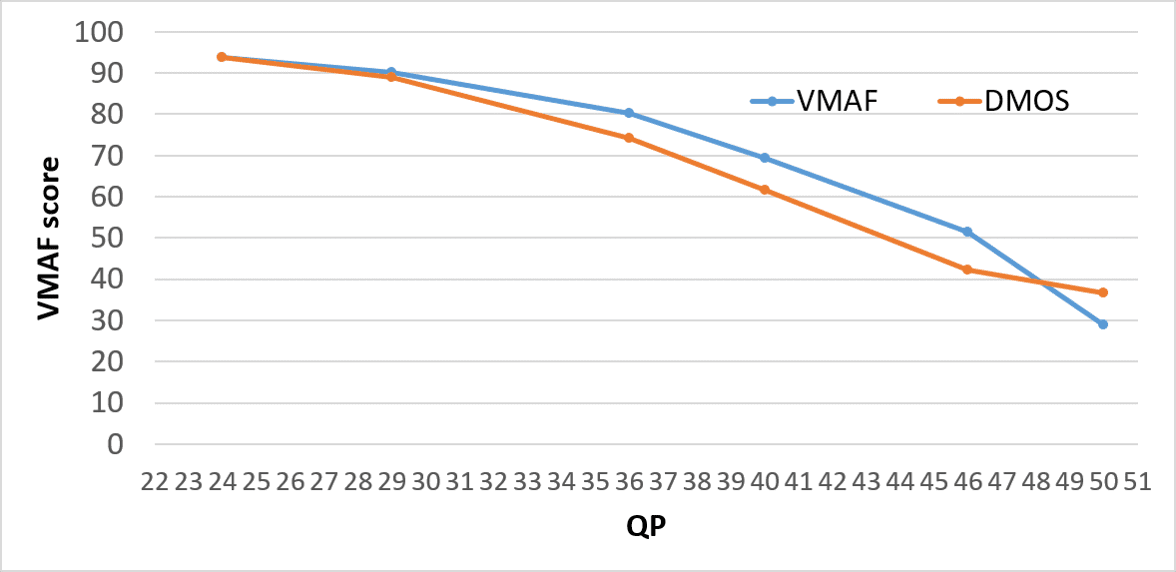}
    }
    \\
    
    \subfloat[\textit{FemaleBasket}.]
    {
        \includegraphics[width=0.4\textwidth]{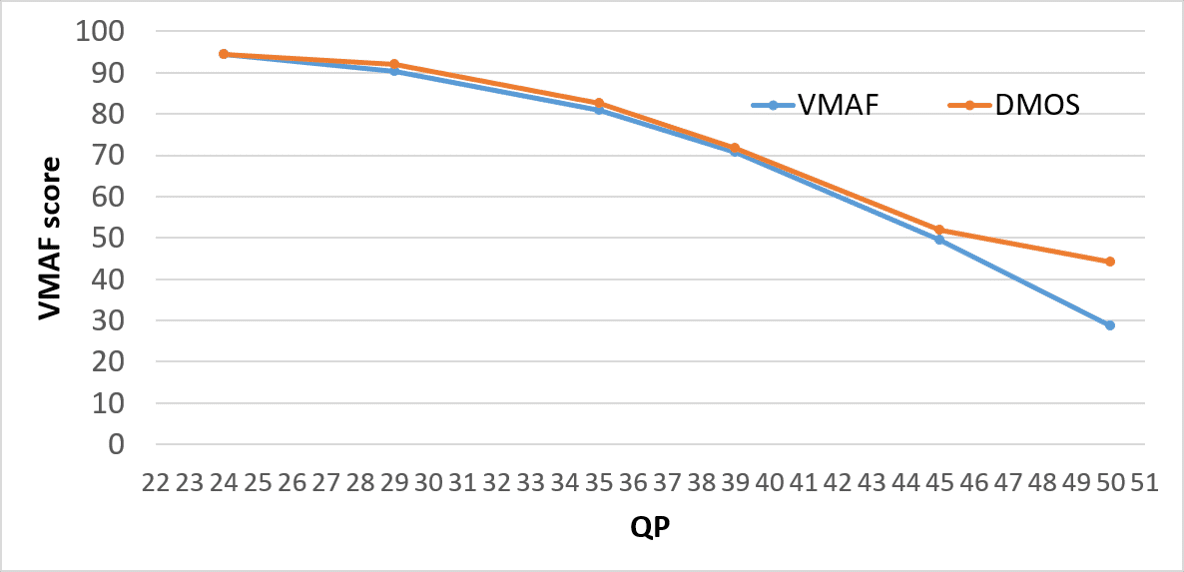}
    }
     \subfloat[\textit{Happyland}.]
    {
        \includegraphics[width=0.4\textwidth]{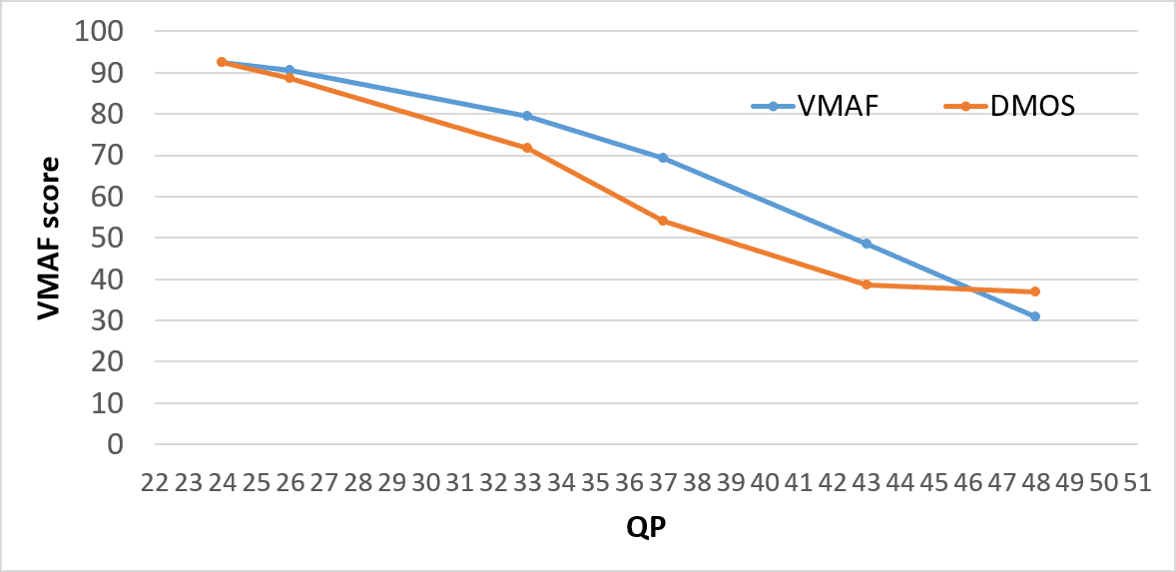}
    }
    \\
    
    \subfloat[\textit{Lions}.]
    {
        \includegraphics[width=0.4\textwidth]{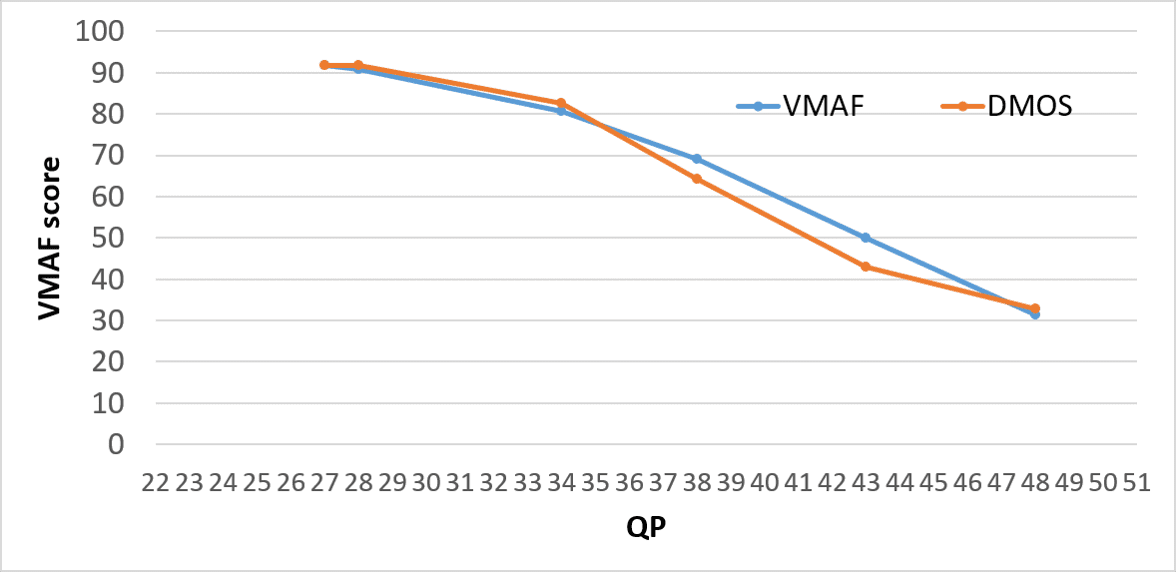}
    }
    \subfloat[\textit{Sunset}.]
    {
        \includegraphics[width=0.4\textwidth]{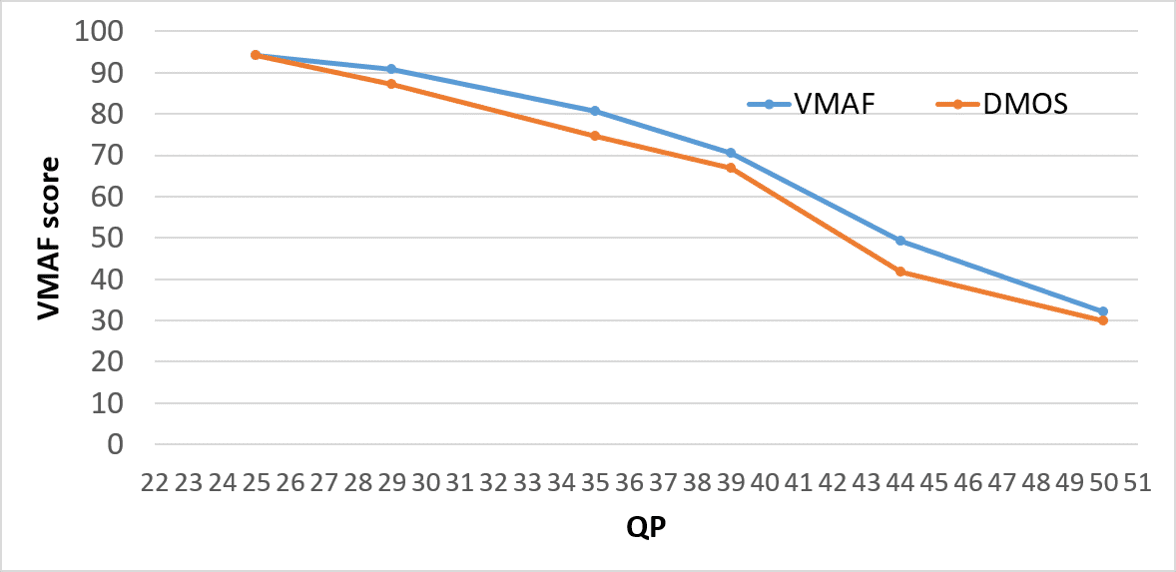}
    }
    \\
    \subfloat[\textit{Waterfall}.]
    {
        \includegraphics[width=0.4\textwidth]{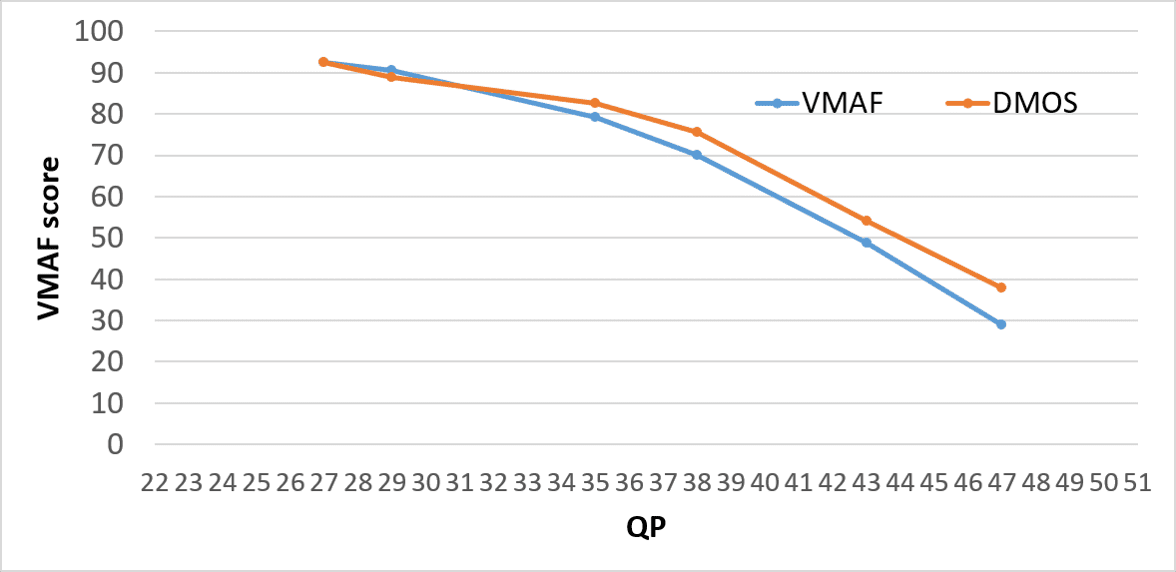}
    }

    \caption{Evolution of the VMAF scores and the normalized DMOS with the QP value for each content.}
    \label{fig:VMAF_DMOS}
\end{figure*}

\begin{table*}[t]
\renewcommand{\arraystretch}{1.3}
\caption{Pearson correlation and RMSE between VMAF and DMOS for all contents\label{tab:pearson}}
\centering
\begin{tabular}{|c|c|c|c|c|}
\hline
\multirow{2}{*}{\textbf{CONTENT}} & \textbf{PEARSON} & \textbf{PEARSON} & \textbf{RMSE} & \textbf{RMSE}\\
& (QB, QC, QD, QE, QF) & (QB, QC, QD, QE) & (QB, QC, QD, QE, QF) & (QB, QC, QD, QE) \\
\hline
\textit{AbandonedBuilding}                      & 0.995                              & 0.997   & 3.433   & 1.983                        \\ \hline
\textit{Alaska}                                 & 0.992                              & 0.994   & 5.661   & 2.488                \\ \hline
\textit{Beach}                                  & 0.992                              & 0.991   & 4.213   & 2.470            \\ \hline
\textit{CaribbeanVacation}                      & 0.961                              & 0.997   & 6.982   & 6.787                \\ \hline
\textit{FemaleBasket}                           & 0.984                              &
1.000   & 7.097   &  1.764            \\ \hline
\textit{Happyland}                              & 0.940                              & 0.979   & 9.338   & 9.991              \\ \hline
\textit{Lions}                                  & 0.987                              & 0.997   & 4.029   &  4.446             \\ \hline
\textit{Sunset}                                 & 0.996                              & 0.998   & 5.016   &  5.490          \\ \hline
\textit{Waterfall}                              & 0.996                              & 0.990   & 5.511   &  4.295       \\ \hline
\multicolumn{1}{|c|}{\textbf{AVERAGE}} & \textbf{0.983}                            & \textbf{0.994}      & \textbf{5.698}       & \textbf{4.413}\\ \hline
\end{tabular}%
\end{table*}

\subsection{Experimental results\label{sec:subjresults}}

Given that the ACR-HR method was implemented, both the Mean Opinion Score (MOS) and the Differential Quality Score~(DMOS) are computed from the evaluations provided by the subjects. The final scores per content and quality are depicted in Figure~\ref{fig:mos} and Figure~\ref{fig:dmos}, respectively. Moreover, 95\% Confidence Intervals~(CI) are included to properly measure the agreement between subjects~\cite{streijl2016mean}, according to the Recommendation ITU-R BT.500-13~\cite{bt2012methodology}.

In Figure~\ref{fig:VMAF_DMOS}, the VMAF scores and the normalized DMOS are presented for each content. So, we can properly and easily compare the curves obtained for each of the measurements. To compute the normalized values, we have considered that the normalized DMOS associated with the reference clip equals the specific VMAF score of this sequence, and the rest of the values are calculated from it. In this way, we completely remove all external influences in the quality perceived by users. It is worth mentioning that the absence of raw video sources in our test material influences our analysis in terms of the choice of the reference sequence for the subjective assessment and, consequently, the DMOS normalization. However, the alternative of acquiring a new specific database of raw video sources, with its associated problematic acquisition and stitching processes, is beyond the scope of this work.

Through the comparison of the VMAF and DMOS curves for the different contents, we can study the performance of the VMAF metric for omnidirectional content. We can see that the shape of the curves is very similar and the gap between both is quite small. Therefore, we can conclude that the subjective rates obtained in our experiment fit the VMAF scores to a great extent for almost the whole range of qualities. Only for "Happyland" and, more moderately, "CaribbeanVacation", we can really notice a greater gap between the VMAF and DMOS curves.

Nevertheless, we can see that there is a deviation of the DMOS curves with respect to the VMAF curves in the lowest range of qualities (high QP values). The most plausible reason to that is that the perceived video quality goes into a saturation region. That is, users statistically barely perceive any differences between sequences encoded with very high QP value. It is caused by artifacts that appear and are annoying to the user, making much more difficult for him/her to discern between such distorted contents. This saturation effect is further boosted by the characteristics of the HMD. In addition, this effect is also justified considering the computation of VMAF. The CIs associated with the VMAF score are notably higher for low qualities, decreasing the reliability of the results.

To validate these findings, we have computed the Pearson's Linear Correlation Coefficient (PLCC) and the Root Mean Square Error (RMSE) between the VMAF and DMOS values. These results are included in Table~\ref{tab:pearson}. Both measures, PLCC and RMSE, are obtained for qualities ranging from B to F and also from B to E, due to the deviation commented previously. We can confirm the extremely high correlation between the VMAF scores and the normalized DMOS, which is even higher when the last QP is not considered.

Therefore, we can assure that VMAF works properly with 360VR content with homogeneous encoding, providing remarkably good results with no specific training focused on this type of content.

\section{Conclusions\label{sec:conclusions}}
We have presented an exhaustive study on the feasibility of directly applying the original VMAF metric to assess the quality of omnidirectional contents watched using an HMD. Based on the assumption that VMAF scores decrease monotonically with the QP, due to the effect of this encoding parameter in the resulting sequence, we have carried out an experiment consisting of two main steps. First, we have used the original implementation to obtain the VMAF score of a number of 360VR sequences encoded with constant QP in the whole range of possible values so as to capture how it varies with the encoding parameter. Secondly, we have validated the obtained VMAF scores through a subjective assessment. We have done so by creating a second curve per content from a finite number of scores corresponding to several operating points, which have been selected sufficiently spaced. These values are the normalized DMOS obtained in the subjective tests for the subset of input sequences encoded for the specific QP anchor points. The minimum divergence of the two curves in most cases allows us to conclude that VMAF works sufficiently correctly with this homogeneous 360VR content, without performing any particular adjustments to prepare the metric accordingly. However, one can avoid the creation of a specific dataset with rich 360VR content of an acceptable quality and retraining the machine learning algorithm to obtain an omnidirectional-content-aware VMAF metric, which, additionally, would be very heavy in terms of computing and time resources.

% or


%\appendices
%% you can choose not to have a title for an appendix
%% if you want by leaving the argument blank
%\section{}
%% use section* for acknowledgment
%\section*{Acknowledgment}

\ifCLASSOPTIONcaptionsoff
  \newpage
\fi



%\bibliographystyle{IEEEtran}
%\begin{thebibliography}{1}
%\end{thebibliography}

\bibliographystyle{IEEEtran}
\bibliography{bibliografia}

%\end{IEEEbiography}
%% insert where needed to balance the two columns on the last page with
%%\newpage


%\vfill


\end{document}